\documentclass[jkps,preprint,fleqn,showpacs,showkeys]{revtex4}
\usepackage{graphicx}

\begin{document}
\title[]{Dirac Coupled-channel Analyses of Polarized Proton Scatterings to the 2$^+$ Gamma Vibrational
Band in $^{24}$Mg and $^{26}$Mg}
	
\author{Sugie \surname{Shim}}
\email{shim@kongju.ac.kr}
\thanks{Fax: +82-41-850-8489}
\affiliation{Department of Physics, Kongju National University, Gongju 32588, Republic of Korea}
\received{2016}

\begin{abstract}
Dirac coupled channel calculations are performed phenomenologically for the high-lying excited states that belong to the 2$^+$ gamma vibrational
band at the 800-MeV polarized proton inelastic scatterings from the s-d shell nuclei, $^{24}$Mg and $^{26}$Mg. Optical potential model is used and
scalar and time-like vector potentials are considered as direct potentials. First-order vibrational
collective models are used to obtain the transition optical potentials in order to accommodate the high-lying excited vibrational collective states. The complicated
Dirac coupled channel equations are solved  phenomenologically to reproduce the
differential cross section and analyzing power data by varying the optical potential and deformation parameters.
 It is found that the relativistic Dirac coupled channel calculation could describe the high-lying excited states of the 2$^+$ gamma vibrational band at the 800-MeV polarized proton inelastic scatterings from s-d shell nuclei $^{24}$Mg and $^{26}$Mg reasonably well, showing better agreement with the experimental data compared to the results obtained from the nonrelativistic calculations. Calculated deformation parameters for the excited states are analyzed and compared with those of nonrelativistic calculations.
\end{abstract}



\maketitle

\section{INTRODUCTION}
Relativistic Dirac approaches based on the Dirac equation have been very successful in treating nuclear reactions \cite{1,2,3}.
 Because the Dirac analyses have proven to be very successful in describing the intermediate-energy proton elastic scatterings from the spherically symmetric nuclei and a few deformed nuclei \cite{3,4,5,6}, the relativistic approaches based on the Dirac equation are expanded to inelastic scatterings and have shown considerable improvements compared to the conventional nonrelativistic calculations based on the Schr\"{o}dinger equation \cite{7,8,9,10}. The relativistic effect is taken into account at the level of the kinematics in the nonrelativistic calculations based on the Schr\"{o}dinger equation, because it is no longer negligible for the intermediate-energy proton scattering. But the relativistic kinematics correction seems not to be enough, so the fully relativistic treatment based on the use of the Dirac equation seems to be required for the description of intermediate-energy proton scattering.
 It should be noted that one of the merit of the relativistic approach based on the Dirac equation instead of using the nonrelativistic approach based on the Schr\"{o}dinger equation is that the spin-orbit potential appears naturally in the Dirac approach when the Dirac equation is reduced to a Schr\"{o}dinger-like second-order differential equation, while the spin-orbit potential should be inserted by hand in the nonrelativistic Schr\"{o}dinger approach in order to describe the intermediate energy nucleon scattering from the nucleus.

In this work we performed the Dirac coupled channel analyses for the high-lying excited states of s-d shell nuclei $^{24}$Mg and $^{26}$Mg that belong to the $K^\pi = 2^+ $  gamma vibrational
band at polarized 800-MeV proton inelastic scattering. Dirac phenomenological optical potentials are used, employing the S-V optical potential model \cite{1} where only scalar and time-like vector potentials are considered. The Woods-Saxon shape is used for the geometry of the direct optical potentials, assuming the shape of the potential follows the shape of the nuclear density.
In order to accommodate the  collective  motion of the excited deformed nucleus considering the high-lying excited states of 2$^+$ gamma vibrational
band, the first-order vibrational collective model is used to obtain the transition optical potentials \cite{3,4}.
The complicated Dirac coupled-channel equations are solved phenomenologically to reproduce the experimental data by varying the optical potential and deformation parameters, using a computer program called ECIS \cite{11}. The effect of the multistep process is investigated by
including the channel coupling between two adjacent excited states in addition to the couplings
between the ground state and the excited states.  The calculated results are analyzed and compared with the experimental data and the results obtained from the nonrelativistic approaches.

\section{Theory and Results}

Relativistic Dirac coupled channel analyses are performed phenomenologically for the high-lying excited states of the 2$^+$ gamma vibrational
band at $^{24}$Mg (p,p$'$) and $^{26}$Mg (p,p$'$) by using an optical potential model and the first-order vibrational collective model.
Because $^{24}$Mg and $^{26}$Mg are the spin-0 nuclei, only scalar, time-like vector and tensor
optical potentials survive \cite{11,12}, as in spherically symmetric nuclei \cite{13}, hence the relevant Dirac equation for the elastic scattering from the nucleus is given as

\begin{equation}
[\alpha \cdot p + \beta ( m + U_S ) - ( E - U_0- V_c )
 + i \alpha \cdot  \hat{r} \beta U_T ] \Psi(r) = 0
\label{e1}
\end{equation}

Here, $U_S$ is a scalar potential, $U_0 $ is a time-like vector potential,  $U_T$ is a tensor potential, and $V_c $ is the Coulomb potential.
However, depending on the  model assumed, pseudo-scalar and
axial-vector potentials may also be present in the equation when we consider inelastic
scattering. We assume, in the collective model approach used in this work, that the appropriate transition potentials can be obtained by deforming the direct potentials that describe the elastic
channel reasonably well \cite{14,15}.
The shapes of the deformed potentials are assumed to follow the shape of the deformed nuclear densities and the transition potentials are obtained by assuming that they are proportional to the first-order derivatives of the diagonal potentials \cite{3,4}. As direct potentials, the scalar and the time-like vector potentials are used in the calculation. Tensor potentials are always present due to the interaction of the anomalous magnetic moment of the projectile with the charge distribution of the target. However, the tensor potentials are neglected in this calculation because they have been found to be always very small compared to scalar or vector potentials \cite{3}.
The evidence that the large scalar and vector fields of Dirac phenomenology may be related to quark degrees of freedom in the nucleon can be found in the work of Cohen ${\it et}$ ${\it al.}$ \cite{16}.
In the vibrational model of ECIS, the deformation of the nuclear surface is written using the Legendre polynomial expansion method as
\begin{equation}
R(\theta, \phi ) = R_0 ( 1+ \sum_{\lambda \mu } \beta_\lambda Y^* _{\lambda \mu } (\theta, \phi ) ),
\label{e2}
\end{equation}
with $R_0$ being the radius at equilibrium, $\beta$ a deformation parameter and $\lambda$ the multipolarity.
The transition potentials for the channel coupling are given by

\begin{equation}
U_i ^\lambda = \frac{\beta^i _\lambda R_i }{(2\lambda +1)^{1/2}} \frac{dU_i (r)}{dr } Y^* _{\lambda \mu} (\Omega)
\label{e2}
\end{equation}
where the subscript $i$  refers to the real and the imaginary scalar or vector potential, $R$ is the radius parameter of the Woods-Saxon shape.  By assuming that the real and the imaginary deformation parameter $\beta^i _\lambda$ are  equal for the given  potential type, two deformation parameters, $\beta_S$ and $\beta_V$, are determined for each excited state.

In order to obtain the Dirac coupled channel equations, we expand the Dirac wavefunction using upper and lower component and substitute them into the Dirac equation. After some calculation\cite{3}, we can obtain the coupled equations for the radial upper component, $g_j$, and lower component $f_j$ of the Dirac spinors as follows.
\begin{eqnarray}
\left[\frac{d}{dr} + \frac{\chi_j}{r}\right]g_j & - &
[E-\epsilon_j + m + U_S^0-U_0^0]f_j + (U_T^0-\frac{\nu }{2m}
\frac{\partial V_c}{\partial r})g_j \nonumber \\
&=&\sum_{\lambda j'} P^\lambda_{jj'}[(U_S^\lambda-U_0^\lambda)
(sgn\chi_j)(sgn\chi_{j'})f_{j'}- U_T^\lambda g_{j'}] \\
\left[\frac{d}{dr} - \frac{\chi_j}{r}\right]f_j & + &
[E-\epsilon_j - m - U_S^0 - U_0^0]g_j-(U_T^0-\frac{\nu }{2m}
\frac{\partial V_c}{\partial r})f_j \nonumber \\
&=&\sum_{\lambda j'} P^\lambda_{jj'}[(U_S^\lambda + U_0^\lambda)
g_{j'}+ U_T^\lambda f_{j'}]
\end{eqnarray}

Here, we separate the elastic optical potentials, $U_i^0$, the scalar and
time-like vector potentials from
transition potentials, $U_i^{\lambda}$, and the subscript $\lambda$ refers to the multipole order of the
particular transition potential. $\chi$ is the quantum number related to the projectile, $E$ is the total energy of the system, $\epsilon_j$ is the nuclear energy eigenvalue, $\nu$ is
the anomalous magnetic moment of the projectile, and
 $P^\lambda_{jj'}$ is the geometric coefficient related to $j$ and $j'$ channel of multipole order \cite{3}.
 These complicated Dirac coupled channel equations are solved numerically to calculate the scattering observables such as differential cross section and analyzing power by using
a computer code, ECIS \cite{11} written by
J. Raynal, which employs the sequential iteration method.

The experimental data for the differential cross sections and analyzing powers are obtained from Ref. 17 for the 800-MeV polarized proton inelastic scatterings from the $^{24}$Mg and $^{26}$Mg nuclei. The high-lying excited states of the 2$^+$ gamma vibrational band, the second $2^+$(4.24 MeV) and the second $4^+$(6.01 MeV) states for the $^{24}$Mg (p,p$'$), the second $2^+$(2,94 MeV) and the second $4^+$(4.90 MeV) states for the $^{26}$Mg (p,p$'$) are considered and assumed to be collective vibrational states in the calculation. One can assign the $3^+$ state, which is an unnatural parity state, to the gamma band for both nuclei, but they are neglected in the calculation because they have large error bars and pretty irregular shape \cite{17}.
First, the 12 parameters of the diagonal scalar and vector potentials in the Woods-Saxon shapes are determined by fitting the experimental data for the elastic scattering from $^{24}$Mg.
The Dirac equations are solved numerically to obtain the best fitting parameters to the experimental data by using the minimum $\chi^2$ method.
 Calculated results for the 800-MeV p +  $^{24}$Mg elastic scattering are shown as dash-dot-dot lines in Figure 1 and it is found that the observable elastic experimental data are reproduced quite well.

\begin{figure}[!]
\includegraphics[width=10cm]{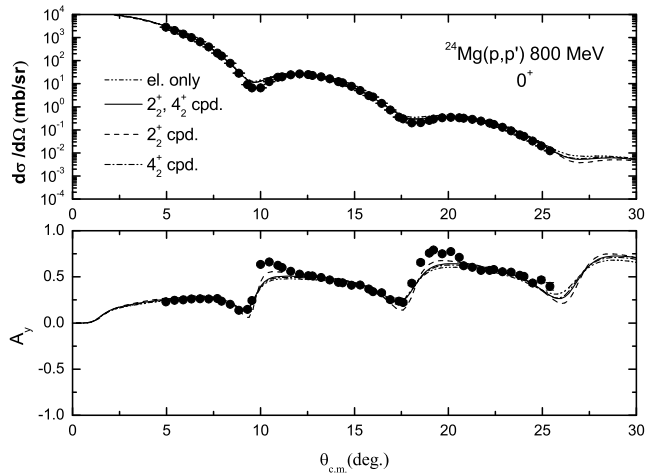}
\caption[0]{Differential cross section and analyzing power of the ground state for 800-MeV p +  $^{24}$Mg scattering. The dash-dot-dot line, dash-dot, dashed and solid lines represent the results of the Dirac phenomenological calculation where only the ground state is considered, where the ground and the $4_2^+$ states are coupled, where the ground, the $2_2^+$ states are coupled, and where the ground, the $2_2^+$ and the $4_2^+$ states are coupled, respectively.}
\end{figure}

\begin{table}[h]
\caption[0]{Calculated phenomenological optical potential parameters of the Woods-Saxon shape for 800-MeV proton elastic scattering from $^{24}$Mg.}
\begin{ruledtabular}
\begin{tabular}{cccccccc}
   Potential      &   Strength (MeV)   & Radius (fm)  & Diffusiveness (fm)  ~ \\
   \hline
 Scalar  & -310.3     & 2.661 & 0.8014        ~ \\
 real    &      &     &            ~ \\ \hline
 Scalar  & 21.70     & 2.299 & 0.7990        ~ \\
 imaginary    &      &     &            ~ \\ \hline
 Vector  & 118.2    & 2.783 & 0.8192         ~ \\
 real    &      &     &           ~ \\ \hline
 Vector  & -82.28     & 2.631 & 0.6437       ~ \\
 imaginary  &      &     &             ~ \\
\end{tabular}
\end{ruledtabular}
\end{table}

Calculated optical potential parameters of the Woods-Saxon shape for the 800-MeV proton elastic scatterings from $^{24}$Mg are shown in Table 1.
 It is confirmed that the real scalar potentials and the imaginary vector potentials turn out to be large and negative, and that the imaginary scalar potentials and the real vector potentials turn out to be large and positive, showing the same pattern as for spherically-symmetric nuclei \cite{3}.

Next, six-parameter searches are performed including the second $2^+$ ($2_2^+$) state which is the lowest excited state of the 2$^+$ gamma vibrational band in addition to the ground state, starting from the obtained 12 parameters for the direct optical potentials. Here, six parameter determine the two deformation parameters, $\beta_S$ and $\beta_V$, of the excited state and the four potential strengths; the scalar real and imaginary potential strengths and the vector real and imaginary potential strengths, keeping the potential geometry unchanged. Here, we varied the optical potential strengths obtained by fitting the elastic scattering data in the elastic scattering calculation because the channel coupling of the excited states to the ground state should be included in the inelastic scattering calculation.
As a next step, another six-parameter searches are performed by including the second $4^+$ ($4_2^+$) excited states in addition to the ground state.
Finally, an eight-parameter search is performed by considering all three states, the ground, the $2_2^+$ and the $4_2^+$ states, together in the calculation, and the results are compared with those of the calculation where only the ground and one excited states are coupled, in order to investigate the effect of the channel coupling between the excited states.

\begin{figure}[!]
\includegraphics[width=10cm]{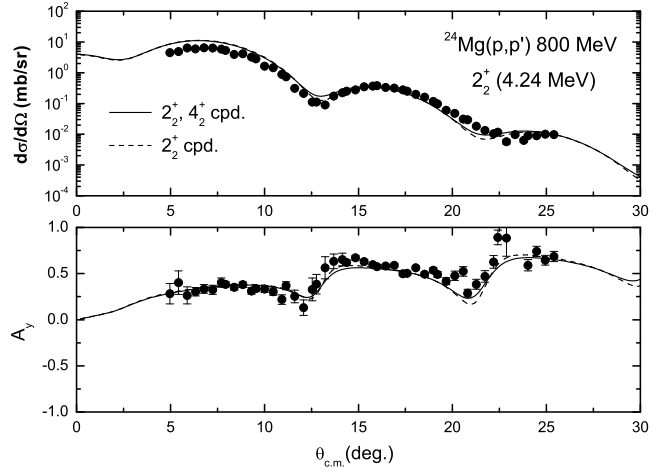}
\caption[0]{Differential cross section and analyzing power of the $2_2^+ $ state for 800-MeV p +  $^{24}$Mg scattering. The dashed, and solid lines represent the results of the Dirac coupled channel calculation where the ground and the $2_2^+$ states are coupled, where the ground, the $2_2^+$ and the $4_2^+$ states of the 2$^+$ gamma vibrational band are coupled, respectively.}
\end{figure}

\begin{figure}[!]
\includegraphics[width=10cm]{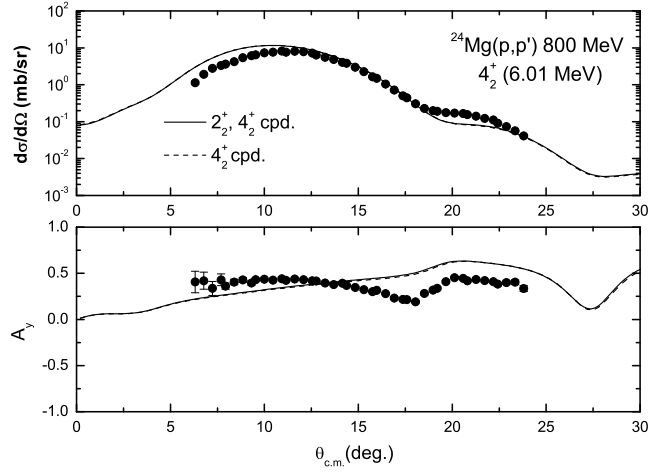}
\caption[0]{Differential  cross section and analyzing power of the $4_2^+ $ state for 800-MeV p +  $^{24}$Mg scattering. The dashed, and solid lines represent the results of the Dirac coupled channel calculation where the ground and the $4_2^+$ states are coupled, where the ground, the $2_2^+$ and the $4_2^+$ states of the 2$^+$ gamma vibrational band are coupled, respectively.}
\end{figure}

The results of the Dirac coupled channel calculations for the ground state for 800 MeV p +  $^{24}$Mg scattering are given in Figure 1. Most of them are shown to reproduce the elastic experimental data pretty well even though the improvement due to the channel coupling effect at large angles of the differential cross section \cite{3,8} is not clearly shown in this case, because the data exist only up to angles of less than 30 degrees.
The calculated observables for the $2_2^+$ state excitation are given in Figure 2. The dashed and the solid line represent the results of the calculation where the ground and the $2_2^+$ states are coupled, where the ground, the $2_2^+$ and the $4_2^+$ are coupled, respectively. The agreements with the experimental data for the $2_2^+$ state are shown to be pretty good for both cases, even though the theoretical lines are slightly out of phase with the data at the position of the second minimum of the diffraction pattern. It is shown that the agreements with the $2_2^+$ experimental data are not changed significantly by adding the $4_2^+$ state excitation in the calculation.
Figure 3 shows the calculated results for the excitation of the $4_2^+$ state. The dashed and the solid line represent the results of the calculation where the ground and the $4_2^+$ states are coupled, where the ground, the $2_2^+$ and the $4_2^+$ are coupled, respectively. The agreements with the experimental data for the $4_2^+$ state in both cases are reasonably good, showing almost the same fit to the experimental data.
Hence it seems that the coupling effects between the excited states of the 2$^+$ gamma vibrational band are insignificant for these cases.
Clearly, better agreements with the experimental data for both excited states are obtained compared to the results obtained from the nonrelativistic calculations \cite{17,18,19}.

The results of the Dirac coupled channel calculations for the ground state for 800-MeV p +  $^{26}$Mg scattering are given in Figure 4. Calculated optical potential parameters of the Woods-Saxon shape for the 800-MeV proton elastic scatterings from $^{26}$Mg are shown in Table 1 of Ref. 20.
Calculated observables of the $2_2^+$ state excitation are given in Figure 5. The dashed and the solid lines represent the results of the calculation where the ground and the $2_2^+$ states are coupled, where the ground, the $2_2^+$ and the $4_2^+$ states are coupled, respectively. Again, the agreements with the experimental data for the $2_2^+$ state are shown to be pretty good for both cases, even though the theoretical lines are slightly out of phase with the data at the position of the second minimum of the diffraction pattern, showing the same pattern as seen in the case of $^{24}$Mg. It is shown that the agreements with the $2_2^+$ experimental data are slightly better in the case where only the $2_2^+$ states are coupled than in the case where the $2_2^+$ and the $4_2^+$ states are coupled.
Figure 6 shows the calculated results for the excitation of the $4_2^+$ state. It is observed that the theoretical lines almost overlap indicating the channel coupling effect between the excited $2_2^+$ and $4_2^+$ states of the 2$^+$ gamma vibrational band is negligible. Hence the direct excitation from the ground state seems dominant for the $4_2^+$ state, as previously suggested in Ref. 17. The agreement with the experimental data for the $4_2^+$ state is reasonably good, showing better agreement with the experimental data compared to the results obtained from the nonrelativistic calculations \cite{17,18}. The discrepancy between the calculated results and $4_2^+$ state data could be explained by the coupling effects with other excited states nearby those are not included in this calculation.

\begin{figure}[!]
\includegraphics[width=10cm]{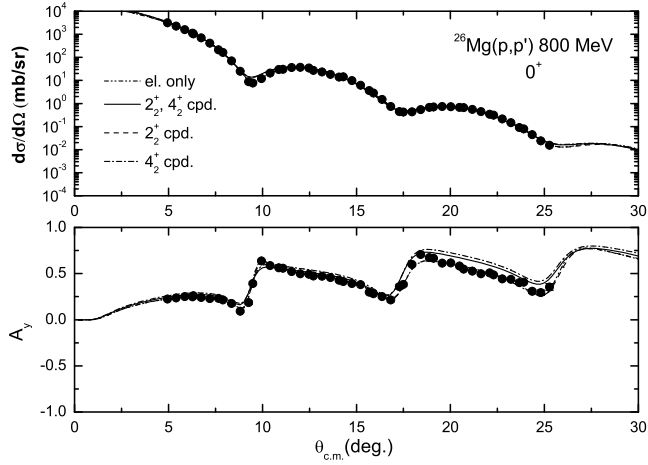}
\caption[0]{Differential cross section and analyzing power of the ground state for 800-MeV p +  $^{26}$Mg scattering. The dash-dot-dot line, dash-dot, dashed and solid lines represent the results of the Dirac phenomenological calculation where only the ground state is considered, where the ground and the $4_2^+$ states are coupled, where the ground, the $2_2^+$ states are coupled, and where the ground, the $2_2^+$ and the $4_2^+$ states are coupled, respectively.}
\end{figure}

\begin{figure}[!]
\includegraphics[width=10cm]{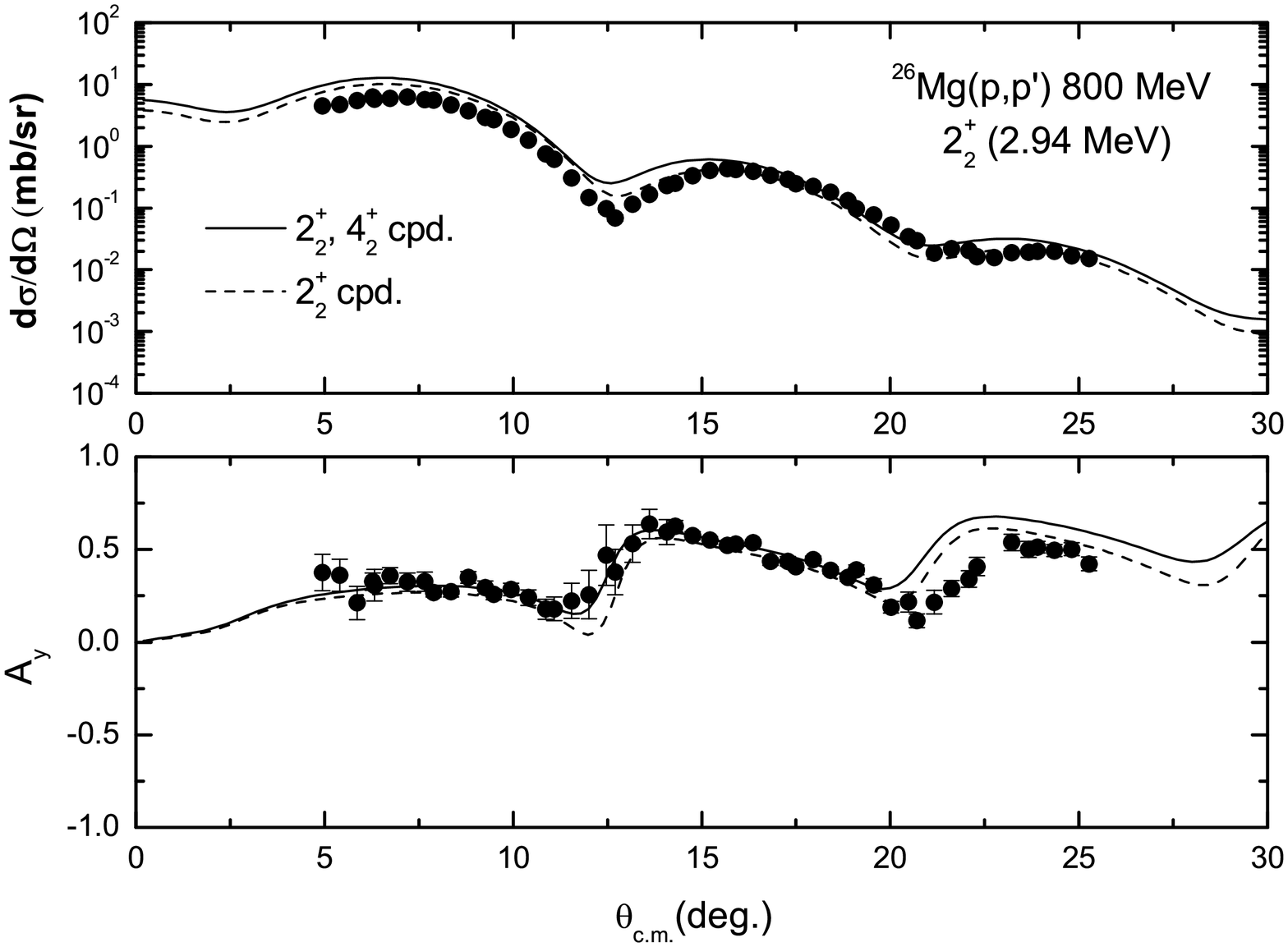}
\caption[0]{Differential  cross section and analyzing power of the $2_2^+ $ state for 800-MeV p +  $^{26}$Mg scattering. The dashed, and solid lines represent the results of the Dirac coupled channel calculation where the ground and the $2_2^+$ states are coupled, where the ground, the $2_2^+$ and the $4_2^+$ states of the 2$^+$ gamma vibrational band are coupled, respectively.}
\end{figure}

\begin{figure}[!]
\includegraphics[width=10cm]{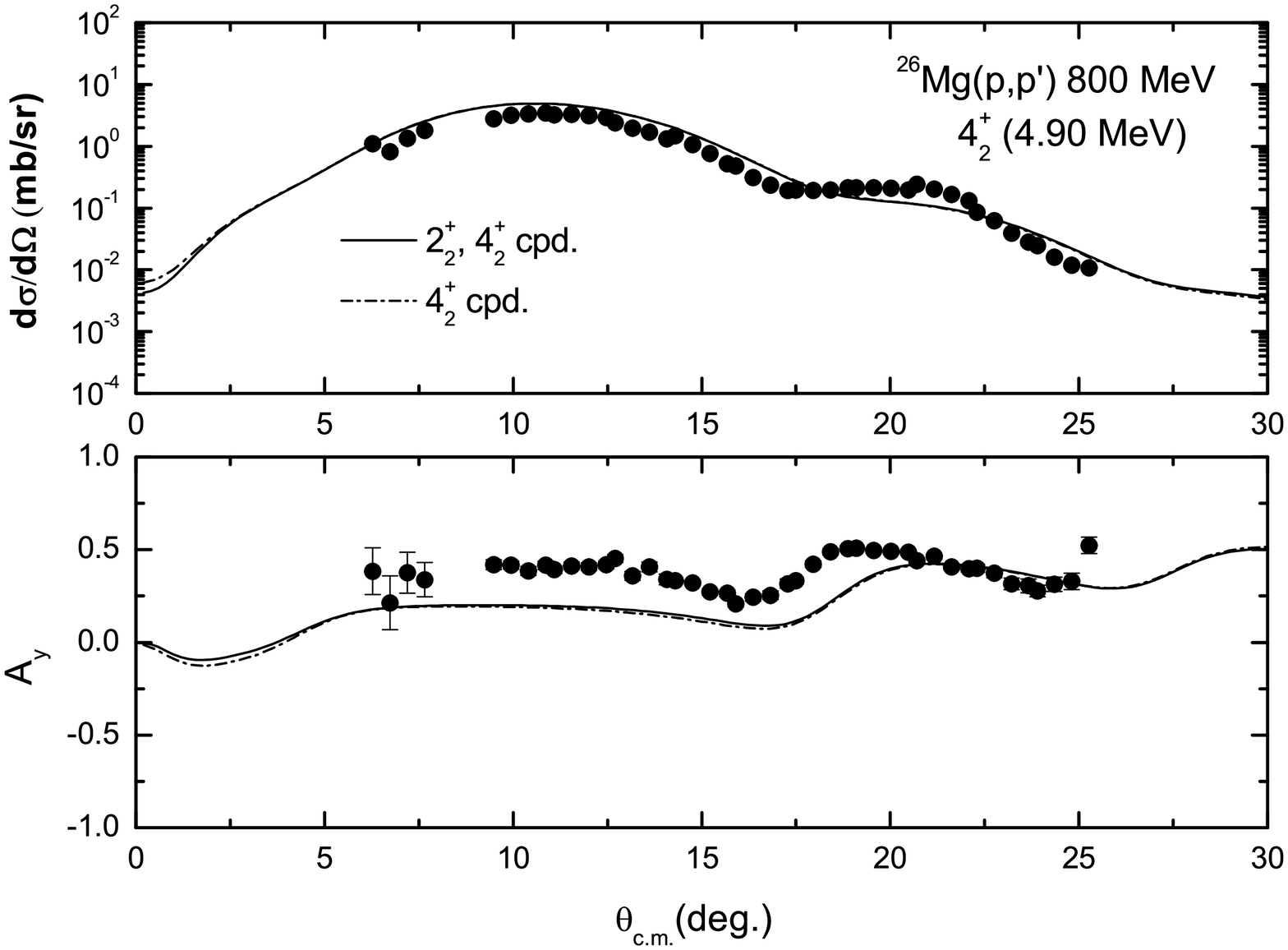}
\caption[0]{Differential  cross section and analyzing power of the $4_2^+ $ state for 800-MeV p +  $^{26}$Mg scattering. The dashed, and solid lines represent the results of the Dirac coupled channel calculation where the ground and the $4_2^+$ states are coupled, where the ground, the $2_2^+$ and the $4_2^+$ states of the 2$^+$ gamma vibrational band are coupled, respectively.}
\end{figure}

\begin{table}[h]
\caption{Comparison of the deformation parameters for the $2_2^+ $ and $4_2^+$ states of the 2$^+$ gamma vibrational band for the 800-MeV proton scatterings from $^{24}$Mg and $^{26}$Mg with those obtained from the nonrelativistic calculations.}
\begin{ruledtabular}
\begin{tabular}{c|ccccccc}
   &Target       &  Energy   &   &   &   ~ \\
   & nuclei      &  (MeV)  & $\beta_S $  & $\beta_V $  & $\beta_{NR} $  ~ \\
   \hline \hline
 $2_2^+ $ state  & $^{24}$Mg & 4.24  &   .204   &  .201   &  $.163^{17}, .187^{19}  $       ~ \\ \cline{2-6}
 & $^{26}$Mg &2.94  &   .174   &  .191   &  $ .142^{17}, .141^{18} $    ~ \\ \hline \hline
 $4_2^+ $ state  & $^{24}$Mg & 6.01  &   .347   &  .274   & $.334^{17}, .333^{19}  $       ~ \\ \cline{2-6}
  & $^{26}$Mg &4.90  &   .102   &  .162   & $ .165^{17},.206^{18}  $    ~ \\
\end{tabular}
\end{ruledtabular}
\end{table}

In Table 2, we show the deformation parameters for the excited states of the 2$^+$ gamma vibrational band in $^{24}$Mg and $^{26}$Mg and compare them with those obtained by using the nonrelativistic calculations. Even though the theoretical bases are quite different, the deformation parameters obtained by using the Dirac phenomenological coupled-channel calculation for the excited states of the 2$^+$ gamma vibrational band in $^{24}$Mg and $^{26}$Mg are shown to  agree pretty well with those obtained by using the nonrelativistic calculations  \cite{17,18,19}. The deformation parameters for the $2_2^+$ and $4_2^+$ states of $^{24}$Mg are found to be larger than those of $^{26}$Mg, indicating that the couplings of the excited states to the ground state are stronger at the $^{24}$Mg (p,p$'$) than at the $^{26}$Mg (p,p$'$), even when the excitation energies for those states are higher at the $^{24}$Mg (p,p$'$) than those at the $^{26}$Mg (p,p$'$).
 The potential strengths are changed to -322.8, 19.68, 123.9 and -80.75 MeV for scalar real and imaginary and vector real and imaginary potentials, respectively, in the $2_2^+$ state and the $4_2^+$ states coupled case for $^{24}$Mg (p,p$'$), -314.1, -21.99, 190.0 and -81.16 MeV in the $2_2^+$ and the $4_2^+$ states coupled case for $^{26}$Mg (p,p$'$).

\section{CONCLUSIONS}

A relativistic Dirac coupled channel calculation using the optical potential model could describe the high-lying excited states of the 2$^+$ gamma vibrational band for the 800-MeV polarized proton inelastic scatterings from the s-d shell nuclei $^{24}$Mg and $^{26}$Mg reasonably well, showing better agreement with the experimental data compared to the results obtained from the nonrelativistic calculation.
 The first-order vibrational collective models are used to describe the high-lying excited collective states of the 2$^+$ gamma vibrational band in the nuclei. The deformation parameters obtained by using the Dirac phenomenological coupled-channel calculation for the high-lying excited states of the 2$^+$ gamma vibrational band in $^{24}$Mg and $^{26}$Mg are found to agree pretty well with the those of the nonrelativistic calculations using the same Woods-Saxon potential shape. The deformation parameters for the $2_2^+$ and the $4_2^+$ states of $^{24}$Mg are turned out to be larger than those of $^{26}$Mg, indicating that the couplings of the excited states to the ground state are stronger at the $^{24}$Mg (p,p$'$) than at the $^{26}$Mg (p,p$'$).
 It is found that the multistep channel-coupling effect seems not important to describe the excited states of the 2$^+$ gamma vibrational band for the proton inelastic scatterings from  $^{24}$Mg and $^{26}$Mg.

\section*{Acknowledgements}

This work was supported by the research grant of the Kongju National University in 2016 and by Basic Science Research Program through the National Research Foundation of Korea(NRF) funded by the Ministry of Education(2016R1D1A1B01014355).


\begin{references}
\bibitem{1} L. G. Arnold, B. C. Clark, R. L. Mercer, and  P. Swandt, Phys. Rev. C {\bf 23}, 1949 (1981).
\bibitem{2} J. A. McNeil, J.  Shepard, and  S. J.  Wallace, Phys.  Rev. Lett  {\bf 50}, 1439 (1983); {\bf 50}, 1443 (1983).
\bibitem{3} S. Shim, Ph. D. Thesis, The Ohio State University 1989: L. Kurth, B. C. Clark, E. D. Cooper, S. Hama, S. Shim, R. L. Mercer, L. Ray, and G. W. Hoffmann, Phys.  Rev. C {\bf 49}, 2086 (1994).
\bibitem{4} S. Shim, B.C. Clark, E.D. Cooper, S. Hama, R.L. Mercer, L. Ray, J. Raynal, and H.S. Sherif, Phys. Rev. C {\bf 42}, 1592 (1990).
\bibitem{5} D. L. Pham and R de Swiniarski, Nuovo Cimento A {\bf 107}, (1994) 1405.
\bibitem{6} R. de Swiniarski, D. L. Pham, and J. Raynal, Z. Phys. A-Hadrons and Nuclei {\bf 343}, 179 (1992).
\bibitem{7} S. Shim {\it et al.}, Int. Jou. of Mod. Phys. E {\bf 21}, 1250098 (2012).
\bibitem{8} S. Shim, M. W. Kim, B. C. Clark, and L. Kurth Kerr, Phys. Rev. C {\bf 59}, 317 (1999).
\bibitem{9} S. Shim {\it et al.}, J. Korean. Phys. Soc. {\bf 51}, 271 (2007): S. Shim et al., J. Korean. Phys. Soc. {\bf 53}, 1146 (2008).
\bibitem{10} J. J. Kelly, Phys. Rev. C {\bf71}, 064610 (2005).
\bibitem{11} J. Raynal, {\it Computing as a Language of Physics}, ICTP International Seminar Course, 281(IAEA, Italy, 1972): J. Raynal, {\it Notes on ECIS94}, Note CEA-N-2772, 1994.
\bibitem{12} C. J. Horowitz and B. D. Serot, Nucl. Phys. A {\bf 368}, 503 (1981).
\bibitem{13} R. J. Furnstahl, C. E. Price, and G.  E. Walker, Phys.  Rev. C  {\bf 36}, 2590 (1987).
\bibitem{14} L. Ray and G. W. Hoffmann, Phys. Rev. C {\bf 31}, 538 (1986).
\bibitem{15} S. Shim and M. W. Kim, J. Korean. Phys. Soc. {\bf 64}, 483 (2014).
\bibitem{16} T. D. Cohen, R. J. Furnstahl, and D. K. Griegel, Phys. Rev. Lett. {\bf 67}, 961 (1991).
\bibitem{17} G. S. Blanpied, B. G. Ritchie, M. L. Barlett, G. W. Hoffmann, J. A. McGill, M. C. Milner, K. W. Jones, S. K. Nanda, R. de Swiniarski, Phys. Rev. C {\bf 37}, 1987 (1988).
\bibitem{18} G. S. Blanpied, N. M. Hintz, G. S. Kyle, M. A. Franey, S. J. Seestrom-Morris, R. K. Owen, J. W. Palm, D. Dehnhard, M. L. Barlett, C. J. Harvey, G. W. Hoffmann, J. A. McGill, R. P. Riljestrand and L. Ray, Phys. Rev. C {\bf 25}, 422 (1982).
\bibitem{19} G. S. Blanpied, N. M. Hintz, G. S. Kyle, J. W. Palm, R. Riljestrand, M. Barlett, C. J. Harvey, G. W. Hoffmann, L. Ray and D. G. Madland, Phys. Rev. C {\bf 20}, 1490 (1979).
\bibitem{20} S. Shim and M. W. Kim, Sae Mulli, {\bf 40}, 545 (2000).
\end{references}
\end{document}